\documentclass[aps,prl,reprint,superscriptaddress,floatfix,amssymb,amsfonts,showpacs]{revtex4-1}

\usepackage{graphicx}
\usepackage{graphicx}
\usepackage{dcolumn}
\usepackage{bm}
\usepackage{lipsum}
\usepackage{amsmath}
\usepackage[mathlines]{lineno}
\newcommand{\forceindent}{\leavevmode{\parindent=1em\indent}}

\begin{document}

%\preprint{APS/123-QED}

\title{Evidence of structural evolution in Sr\textsubscript{2}RhO\textsubscript{4} studied by time-resolved optical reflectivity spectroscopy}

\author{Min-Cheol Lee}
\thanks{Corresponding author}
\email{mclee@lanl.gov}
\thanks{Current affiliation: Center for Integrated Nanotechnologies, Los Alamos National Laboratory, Los Alamos, New Mexico 87545, USA}
\author{Inho Kwak}
\author{Choong H. Kim}
\author{Bumjoo Lee}
\author{Byung Cheol Park}
\author{Junyoung Kwon}
\author{Wonshik Kyung}
\author{Changyoung Kim}
\author{Tae Won Noh}
\affiliation{Center for Correlated Electron Systems (CCES), Institute for Basic Science (IBS), Seoul 08826, Republic of Korea}
\affiliation{Department of Physics and Astronomy, Seoul National University, Seoul 08826, Republic of Korea}
\author{Kyungwan Kim}
\affiliation{Department of Physics, Chungbuk National University, Cheongju, Chungbuk 28644, Republic of Korea}

\date{\today}

\begin{abstract}
We investigate ultrafast dynamics from photoinduced reflectivity of Sr\textsubscript{2}RhO\textsubscript{4} by using femtosecond near-infrared pulses. We observe a clear temperature dependent anomaly in its electronic dynamics which slows down below \textit{T}\textsubscript{S}$\sim$ 160 K. In addition, coherent oscillations of the \textit{A}\textsubscript{1g} symmetric 5.3-THz phonon exhibit a 90$^{\circ}$ shift in its initial phase across \textit{T}\textsubscript{S}, indicating a structural change in octahedral rotation distortions. We propose that octahderal structure in Sr\textsubscript{2}RhO\textsubscript{4} evolves at around \textit{T}\textsubscript{S}, and it can influence on the non-equilibrium dynamics of photoinduced carriers as well as real-time phonon responses.
\end{abstract}

\maketitle

\forceindent{} The interactions between charge, spin, orbital and lattice degrees of freedom are significant to determine quantum phases in correlated electron systems. Ca\textsubscript{2-x}Sr\textsubscript{x}RuO\textsubscript{4} is an crucial example to investigate how the lattice structure governs the ground states ranging from an unconventional superconductor Sr\textsubscript{2}RuO\textsubscript{4} \cite{Maeno1994} to a Mott insulator Ca\textsubscript{2}RuO\textsubscript{4} \cite{Nakatsuji1997}. Sr\textsubscript{2}RuO\textsubscript{4} presents a quasi-two-dimensional layer composed of RuO\textsubscript{6} octahedra without any distortions in bulk \cite{Maeno1994}. Cation substitution with Ca\textsuperscript{2+} ions induces tilting and rotational distortions of the octahedra, which results in Mott gap opening as well as spin/orbital orderings \cite{Nakatsuji2000,Zegkinoglou2005}. In addition, Sr\textsubscript{2}RuO\textsubscript{4} has a distinguished surface state originating from the topmost monolayer with rotational distortions, which hybridize $d$-orbitals and produces an exceptional ferromagnetic order \cite{Matzdorf2000}. It shows that structural deformation especially in the octahedral structures is critical to modify the ground states of correlated materials based on the perovskite structure.\\
\forceindent Layered perovskites Sr\textsubscript{2}RhO\textsubscript{4} and Sr\textsubscript{2}IrO\textsubscript{4} have offered an opportunity to study the effects of octahedral distortions along with the ruthenates. Octahedral rotations about \textit{c}-axis by $\sim$10$^{\circ}$ in Sr\textsubscript{2}RhO\textsubscript{4} and Sr\textsubscript{2}IrO\textsubscript{4} \cite{Subramanian1994} turn their ground states into a clean correleated metallic phase \cite{Perry2006} and an antiferromagnetic insulating phase \cite{Kim2008}, respectively. Intense studies have revealed that Sr\textsubscript{2}IrO\textsubscript{4} is a spin-orbit coupled \textit{J}\textsubscript{eff} = 1/2 magnet, but the magnetic ground state of Sr\textsubscript{2}RhO\textsubscript{4} is still unveiled despite its isostructure to the layered iridate, except a report on a two-dimensional short range antiferromagnetic order \cite{Subramanian1994,Shimura1992}. Moreover, Sr\textsubscript{2}RhO\textsubscript{4} has been rarely investigated in regard to its temperature (\textit{T}) dependence of the octahedral structures except a neutron scattering result \cite{Vogt1996}, although the layered rhodate shows thermal anomalies in the physical quantities of resistivity and magnetic susceptibility \cite{Perry2006,Shimura1992}.\\
\forceindent Here, we performed optical pump-probe spectroscopy on Sr\textsubscript{2}RhO\textsubscript{4} to investigate non-equilibrium dynamics of photoinduced carriers. In time-resolved reflectivity data, we found two different \textit{T}-dependent anomalies: (1) development of slow relaxation dynamics, and (2) variations in amplitude and phase of coherent oscillations of an \textit{A}\textsubscript{1g} symmetric phonon. We observed that the ultrafast dynamics of photoinduced reflectivity slows down below \textit{T}\textsubscript{S}$\sim$160 K. In addition, Sr\textsubscript{2}RhO\textsubscript{4} shows the \textit{T}-dependent coherent oscillations of the $A_{1g}$ symmetric phonon, which corresponds to RuO\textsubscript{6} octahedral rotations in the quasi-two-dimensional plane. The coherent $A_{1g}$ phonon presents clear anomalies across \textit{T}\textsubscript{S} with a huge change in its oscillation-phase by 90$^{\circ}$, and a redshift in frequency below \textit{T}\textsubscript{S}. The unexpected feature in the coherent phonon of Sr\textsubscript{2}RhO\textsubscript{4} suggests structural deformation in the octahedral rotations.\\
\begin{figure}[ht!]
	\includegraphics[width=3.4in]{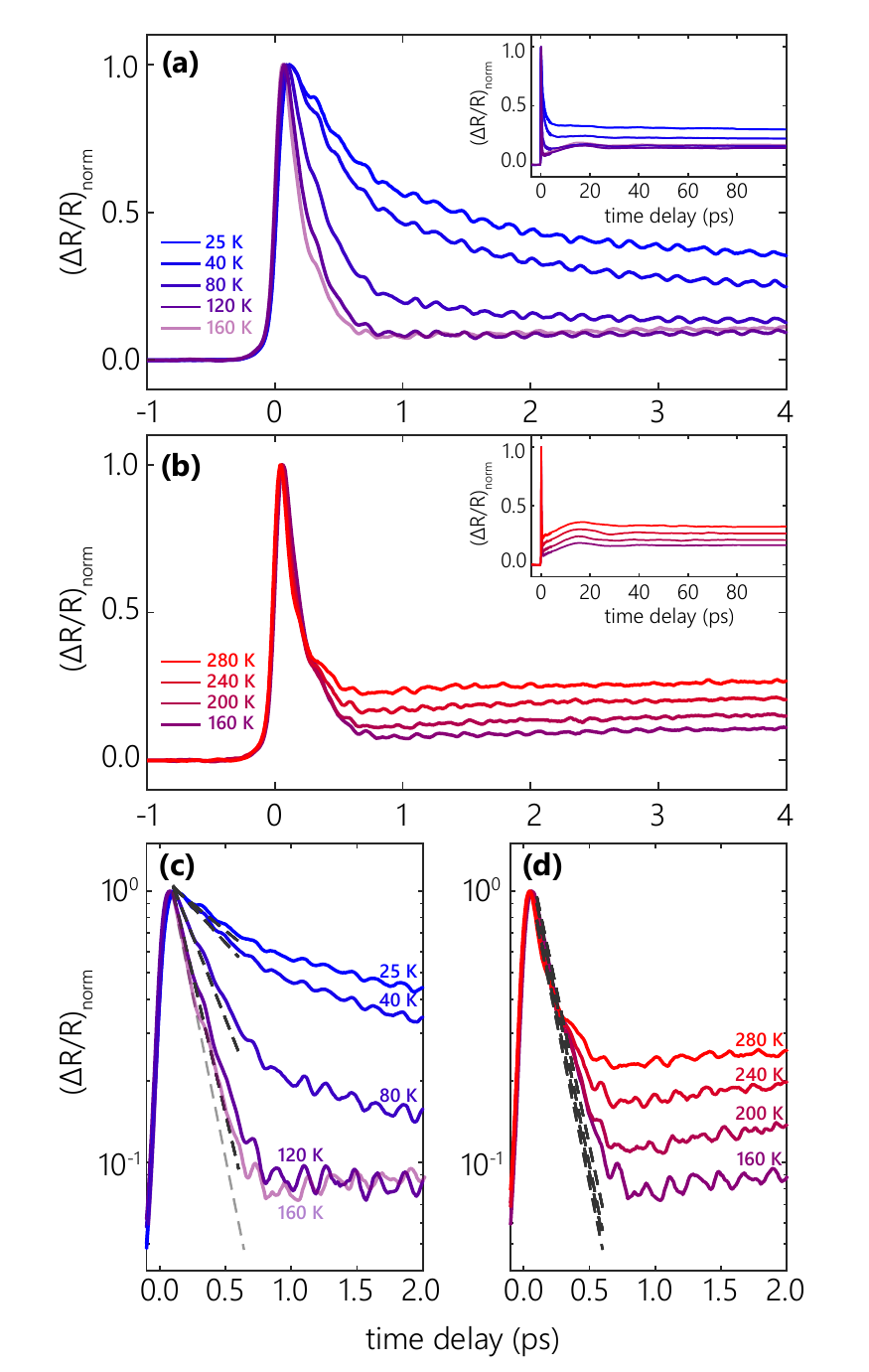}
	\centering
	\caption{Temperature dependent reflectivity changes in Sr\textsubscript{2}RhO\textsubscript{4} stimulated by near-infrared optical pulses at (a) from 25 K to 160 K and (b) from 160 K to 280 K. The data ($\Delta{R}/R$)\textsubscript{norm} were normalized by the maximum peak values. The same data are presented on a logarithmic scale in (c) and (d). The dotted lines are linear fittings to verify the scattering rate of the initial exponential decay.}
	\label{FIG1}
\end{figure}
\forceindent We utilized femtosecond near-infrared pulses to trace real-time reflectivity change by photo-excitations in Sr\textsubscript{2}RhO\textsubscript{4}. Time-resolved reflectivity was measured with 800-nm pulses for pump and probe beams, which were generated by a commercial Ti:Sapphire amplifier system with a 250-kHz repetition rate. The full width half maximum of the spot sizes were set to be 100 and 60 $\mu$m for the pump and probe beams, and we used pump fluences ranging from 20 to 4000 $\mu$J cm\textsuperscript{-2}, and probe fluence of 30 $\mu$J cm\textsuperscript{-2}. The time duration of pump and probe pulses is 30 fs. To make a pump scattering noise minimized, we used light polarizations of pump and probe beams orthogonal to each other.\\
\forceindent Figure 1 shows \textit{T}-dependent time-resolved reflectivity from 25 K to 280 K measured under pump fluence of 85 $\mu$J cm\textsuperscript{-2}. The data ($\Delta{R}/R$)\textsubscript{norm} were normalized by their maximum peak values. The decaying transient of the photoinduced reflectivity is composed of multiple parts; (1) relaxation component and (2) coherent oscillations. Scattering among the hot photoinduced carriers after pumping gives rise to the initial maximum change \cite{Averitt2002,Giannetti2016}. Thereafter, the energy of the hot carriers dissipates into other subsystems like phonons, showing exponential relaxation dynamics. On the top of the overall dynamics, periodic oscillations are observed as shown in Fig. 1. The faster oscillation components with periods of a few hundred femtoseconds are driven by optical phonons \cite{Zeiger1992}, while the slower one with a 15-ps period is generated by a strain wave propagating along $c$-axis \cite{Thomsen1984}.\\
\begin{figure}[t!]
	\includegraphics[width=3.4in]{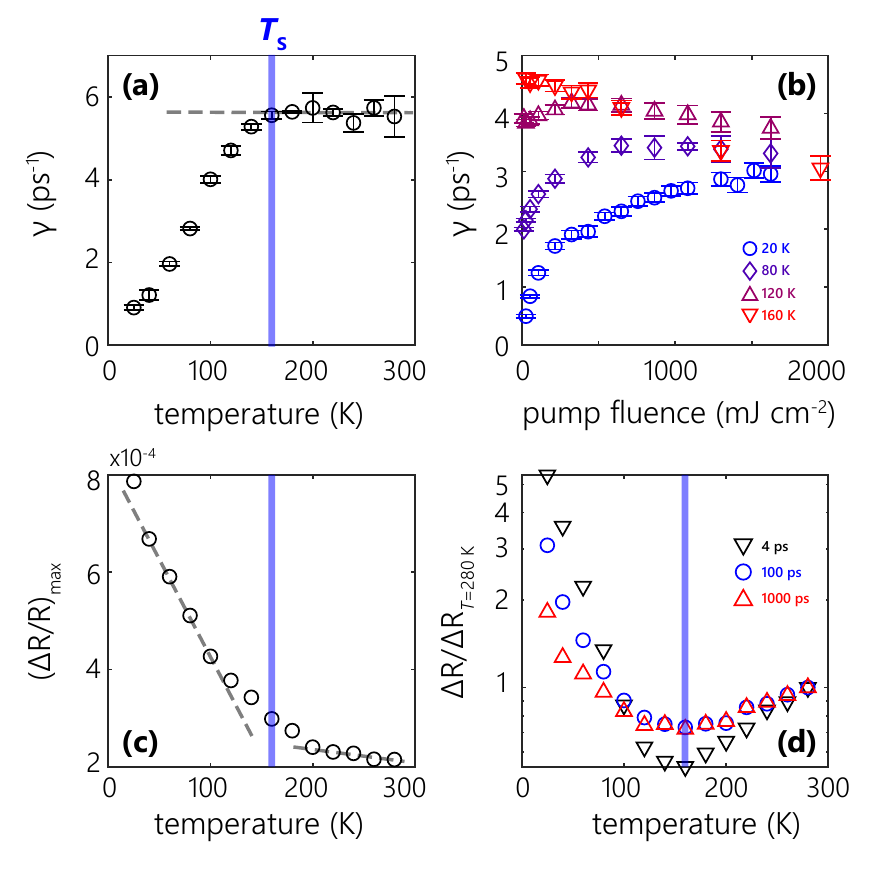}
	\centering
	\caption{Fitting parameters of scattering rate as a function of (a) temperature and (b) pump fluence. Reflectivity change at (c) maximum peak and (d) a few time points at $t$ = 4, 100, 1000 ps. The data of each time point in (d) are normalized by the value at \textit{T} = 280 K.}
	\label{FIG2}
\end{figure}
\forceindent Sr\textsubscript{2}RhO\textsubscript{4} shows clear \textit{T}-dependent variations in its photoinduced reflectivity. Upon cooling, relaxation dynamics becomes slower in low temperatures as shown in Fig. 1(a) and (b). Logarithmic plots in Fig. 1(c) and (d) present the \textit{T}-dependence much clear that the initial relaxation at $t$ $<$ 0.25 ps are almost identical above 160 K, while it gradually slows down below the temperature. To articulate the \textit{T}-dependence, we fitted the reflectivity change on a logarithmic scale to a linear function \cite{scfit}. We only used the data at $t$ $<$ 0.25 ps to extract a scattering rate of an initial exponential decay, and to ignore other contribution from acoustic phonon oscillations or long-lived component. The fitting results are displayed as dotted lines in Fig. 1(c) and 1(d).\\
\forceindent In Fig. 2(a) and (b), we plot the fitting results of scattering rate $\gamma = 1/\tau$, where $\tau$ is the time constant of the initial decay. We found a huge suppression in $\gamma$ at low temperatures, and the \textit{T}-dependent anomaly starts to occur at around \textit{T}$\sim$160 K as temperature decreases (Fig. 2(a)). Such a development of slow relaxation at low temperatures is similar to that of Sr\textsubscript{2}IrO\textsubscript{4} with a energy-gap development below the N\'eel temperature \cite{Hsieh2012}. It has been suggested that the relaxation component originates from electron-hole ($el$-$hl$) recombination process, which gets slower by generation of an energy gap as energy states near the Fermi level disappears \cite{Demsar1999,Gedik2004,Chia2006}.\\
\forceindent For a further analysis, we investigated a pump-fluence dependence of $\gamma$ at various temperatures. Fig. 2(b) shows a scattering rate $\gamma$ which shows a linear dependence to the pump fluence below $\sim$200 $\mu${J} cm$^{-1}$ at \textit{T} $<$ 160 K. It corresponds to the fact that the relaxation rate should be proportional to the number of phtoinduced $el$-$hl$ pairs in the recombination process \cite{Rothwarf1967}. The linearity to the pump fluence remains at \textit{T}= 120 K slightly and disappears at 160 K, which suggests the gap feature starts to evolute at around \textit{T}\textsubscript{S}. Although Sr\textsubscript{2}RhO\textsubscript{4} remains as a correlated metal upon cooling \cite{Perry2006}, angle-resolved photoemission spectroscopy revealed that the a small gap of 10-meV forms below the Fermi level \cite{Baumberger2006}. The energy gap structure was observed at 10 K and it was suggested that the gap is generated by orthorhombicity due to the octahedral rotations. The energy gap value is equivalent to a thermal energy of \textit{T}= 116 K, which could be related to \textit{T}\textsubscript{S} considering errorbar in the photoemission experiment. This indicates that the $el$-$hl$ recombination can be accelerated above \textit{T}\textsubscript{S} as observed by high scattering rate of $\gamma \sim 6$ ps$^{-1}$. We assume that the \textit{T}- and fluence-dependent anomalies in the initial relaxation dynamics can arise from the 10-meV gap, but it still needs a clear explanation how such a gap below the Fermi level can be related to the ultrafast dynamics.\\
\forceindent The raw data of reflectivity transient present additional features of \textit{T}-dependent evolution across \textit{T}\textsubscript{S}. In Fig. 2(c) we plot maximum peak of reflectivity change as a function of temperature. We observed that the peak value shows two different temperature regions with distinct linear \textit{T}-dependences below and above \textit{T}\textsubscript{S}. This can be due to the development of slow relaxation below \textit{T}\textsubscript{S}, as suggested by the decrease of the scattering rate $\gamma$. However, the data do not show a sharp anomaly at \textit{T}\textsubscript{S} but a gradual change in the slope from 120 K to 200 K, proposing that the \textit{T}-dependent evolution can arise in a wide range of temperatures.\\
\forceindent Figure 2(d) shows reflectivity changes $\Delta{R}/R$ at $t$ = 4, 100, and 1000 ps. The data are normalized by the data at 280 K for clear comparison of the \textit{T}-dependence. We found that all of the data present the minimum at \textit{T}= 160 K and distinct \textit{T}-dependences below and above \textit{T}\textsubscript{S} as found by ($\Delta{R}/R$)\textsubscript{max}. The measured time points are far off the time constants of initial relaxations, and therefore $\Delta{R}$($t$)s present an optical response in the quasi-equilibrium state with the same electronic and lattice temperatures; $(R(T+\Delta{T})-R(T))/R(T)$ \cite{Hsieh2012}. From the \textit{T}-dependent anomalies in the reflectivity change dynamics, we suggest crystal structure evolution at around \textit{T}\textsubscript{S}, which can build up the energy gap with a overall change in the electronic structure and can influence the optical response at 1.55 eV.\\
\begin{figure*}[t!]
	\includegraphics*[width=7 in]{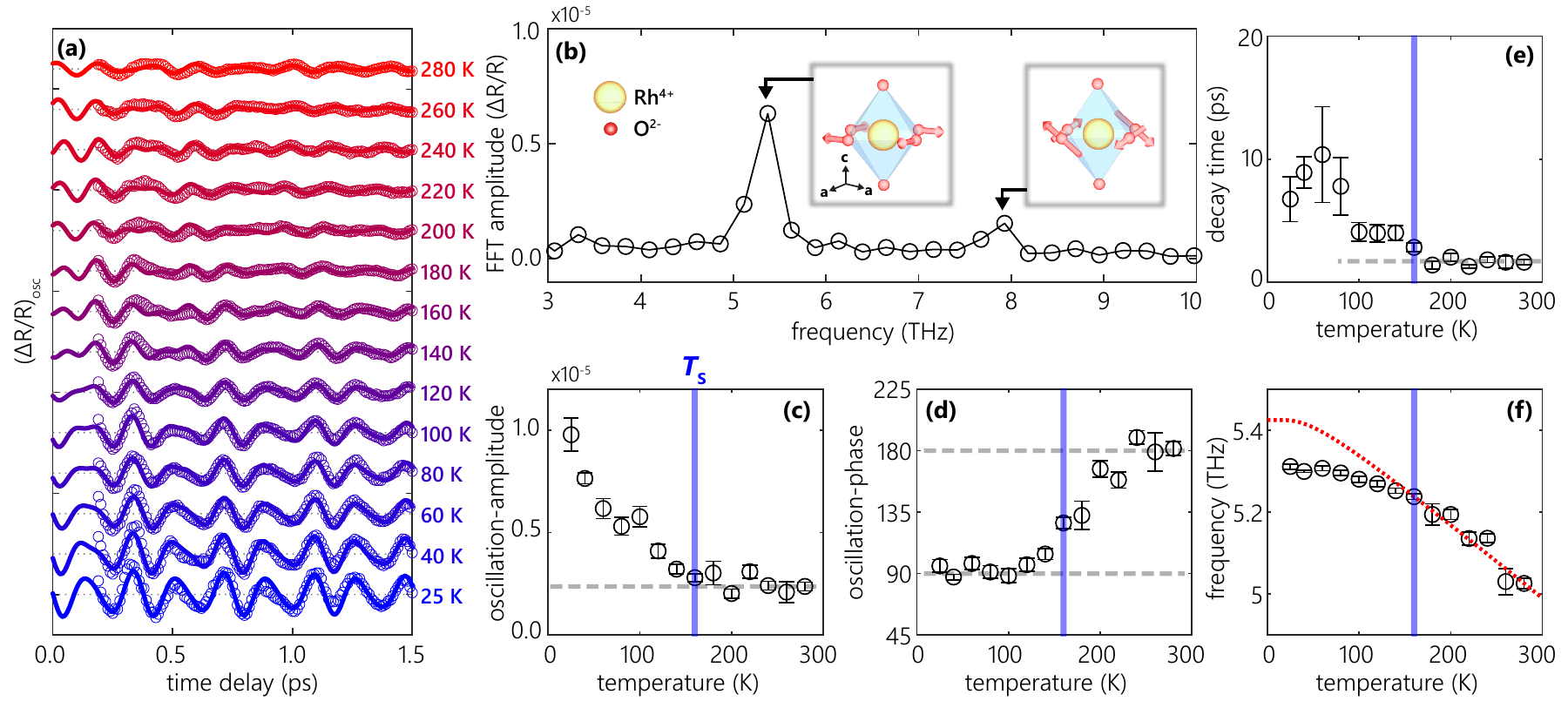}
	\centering
	\caption{(a) Coherent oscillations of photoinduced reflectivity change measured at various temperatures from 25 K to 280 K. We extracted the coherent phonon oscillations by means of bi-exponential decay fitting as metioned in the main text. The data are of $\Delta{R}/R$, which are not normalized by the maximum peaks. (b) Fourier transform of the coherent oscillations in (a). Two resonant oscillations correspond to the $A_{1g}$ symmetric Raman phonons. (c-e) Fitting parameters of the coherent oscillations from the lower $A_{1g}$ phonon. The data are obtained by using a damped harnomic oscillator model. (f) The oscillating component of the lower phonon after subtraction the higher mode oscillations from (a).}
	\label{FIG3}
\end{figure*}
\forceindent To verify the evolution across \textit{T}\textsubscript{S}, we focus on the coherent phonon oscillations -- periodic modulations in optical probing signals driven by Raman active phonons after photoexciations with femtosecond pulses \cite{Zeiger1992,Thomsen1984,Lee2018,Lee2019,Dhar1994,Stevens2002,Garrett1996,Li2013,Riffe2007,Kuznetsov1994}. The real-time phonon observation have offered another scope for crystal structures, as recent studies on Ca\textsubscript{2}RuO\textsubscript{4} revealed that lattice deformation can induce anomalous changes in oscillation-phases, particularly achieved by the ultrafast spectroscopy \cite{Lee2018,Lee2019}. These studies on coherent phonons provided new approaches to investigate octahedral structures in correlated systems.\\
\forceindent Figure 3(a) display the coherently oscillating components by optical phonons ($\Delta{R}/R$)\textsubscript{osc} (open circles), which were extracted from the raw reflectivity transients with the bi-exponential decay model \cite{supple}. Fig. 3(b) shows Fourier transform of the oscillating data, and two resonant modes well correspond to the $A_{1g}$ symmetric Raman phonons as confirmed by our density functional theory (DFT) calculations. While the eigenmode of the 5.3 THz \textit{A}\textsubscript{1g} phonon is solely composed of the octahedral rotations of the in-plane oxygen, the vibrations along \textit{c}-axis are added in the 7.8 THz mode as shown in the inset figure of Fig. 3(b).\\
\forceindent With the values of the resonant frequencies from the Fourier transform, we fit the data by means of a damped harmonic oscillator model:
\begin{equation}
(\Delta{R}/R)_{osc}(t) = -\Sigma_i{C_i}\cos(2\pi{f}_i+\phi_i)\exp(t/\tau_i),
\end{equation}
where ${C_i}, {f_i}, {\phi_i}$, and ${\tau_i}$ present the amplitude, frequency, initial phase, and damping time of the \textit{A}\textsubscript{1g} symmetric phonons. We plot the fitting results as solid line curves in Fig. 3(a), of which the curves well-matched to the experimental data. Figure 3(c-e) show the fitting parameters of the oscillation-amplitude, phase, and decay time of the 5.3 THz mode, which is a major component of the coherent oscillations as shown in Fig. 3(b). The fitting parameters show clear anomalies with a development in the amplitude below \textit{T}\textsubscript{S} (Fig. 3(c)), as well as a large shift in the phase across \textit{T}\textsubscript{S} (Fig. 3(d)). The decay time of the 5.3 THz phonon shows apparent growth below \textit{T}\textsubscript{S} (Fig. 3(e)) while the oscillation frequency gradually increases as temperature decreases (Fig. 3(f)).\\
\forceindent The 90$^\circ$ shift in the oscillation-phase of coherent phonons is not a general case in an absorbing medium. Once a material absorbs a light, equilibrium coordinate of the lattice is assumed to be shifted due to a change in charge density distribution \cite{Zeiger1992}. In other words, the equilibrium position in the photo-excited state ($Q_0^{ex}$) should be moved from the that in the ground state ($Q_0$) (Figure 1 in Ref. \cite{Lee2019}). This discrepancy in the equilibrium locations drives the displacive type force to the lattice generating coherent phonon oscillations with a cosine-type-motion. The displacive motions in the lattice arise as periodic modulation in optical probing signals such as reflectivity with $\Delta{R}_{osc} = (\partial{R}/\partial{Q})\Delta{Q}$. Because Sr\textsubscript{2}RhO\textsubscript{4} absorbs lights with 1.55 eV of the energy of our pumping light at all temperature (not shown), coherent phonon oscillations were expected to be displacive cos-type. However, the 5.3-THz phonon turns into sine-type oscillations below \textit{T}\textsubscript{S}, which has been rarely observed in opaque materials except in the antiferromagnet Ca\textsubscript{2}RuO\textsubscript{4} \cite{Lee2019}.\\
\forceindent The abnormal phase-shift even in the resonant condition can be attributed to a structural modulation along the phonon coordinate. As in the case of Ca\textsubscript{2}RuO\textsubscript{4}, a gradual distortion in the ground state towards the phonon coordinate $\Delta{Q}_{ph}$ can reduce the amplitude of displacive-type oscillations $\Delta{R}_{osc}$ as $\delta{Q}=Q_0^{ex}-Q_0$ becomes zero \cite{Lee2018,Lee2019}. Specifically, as the octahedral structure distorts upon cooling, the \textit{A}\textsubscript{1g} phonon oscillations can turn into sine-type vibrations induced by impulsive stimulated Raman scattering process \cite{Dhar1994,Stevens2002}, which are usually screened out by the displacive mechanism in an opaque material \cite{Stevens2002,Franck1926}. The 90$^\circ$-phase shift of the \textit{A}\textsubscript{1g} phonon oscillations in Ca\textsubscript{2}RuO\textsubscript{4} can be driven by lattice deformation of octahedral tilting distortions with the antiferromagnetic spin ordering \cite{Lee2019}. Accordingly, we argue that structural modulation along the \textit{A}\textsubscript{1g} phonon coordinate in Sr\textsubscript{2}RhO\textsubscript{4} can change the generation mechanism of the coherent phonon. We also found an evidence of the lattice deformation from the previous neutron scattering result \cite{Vogt1996}. As shown in Fig. S2, \textit{T}-dependent rotation angle of RuO\textsubscript{6} indicates a change in a trend of thermal development of the octahedral structure across \textit{T}$\sim$\textit{T}\textsubscript{S}. Additionally, one can note that the phase-shift proceeds from 120 K to 200 K, suggesting a wide \textit{T}-range of the structural evolution as pointed by the maximum values of reflectivity transient (Fig. 2(c)).\\
\forceindent Nevertheless, the stronger phonon oscillation below \textit{T}\textsubscript{S} demands an explanation. The structural distortions in the octahedral rotations account for the suppression in the displacive-type oscillations, but not the larger amplitude of the impulsive-type ones. The strength of the impulsive phonons should be subdued as $\delta{Q}=Q_0^{ex}-Q_0$ becomes zero without any other contribution to the phonon. We found a feature of the interaction to the phonon by the \textit{T}-dependence of oscillation frequency as shown in Fig. 3(f). In general, a gradual redshift as temperature increases originate from the anharnomic effects by phonon-phonon scattering with a model of $\omega(T) = \omega_0 - C[1 + 2/(e^{(\hbar\omega_0/kT)}-1)]$, where $\omega(T)$ is a \textit{T}-dependent phonon frequency and $\omega_0$ is frequency at zero temperature \cite{Balkanski1983}. However, the anharmonic model does not fit to the frequency of 5.3-THz phonon below \textit{T}\textsubscript{S}, while the model almost perfectly fits to the 7.8-THz data (Fig. S3) \cite{supple}. The result indicates an extra interaction to the phonon inducing a 4-meV redshift below \textit{T}\textsubscript{S}.\\
\forceindent As in Sr\textsubscript{2}RhO\textsubscript{4}, Ca\textsubscript{2}RuO\textsubscript{4} presents a redshift in the phonon frequency of the \textit{A}\textsubscript{1g} mode with stronger impulsive amplitudes in the magnetic phase. The spin-phonon coupling originates the softening in the phonon frequency in Ca\textsubscript{2}RuO\textsubscript{4} \cite{Lee2019} as in the cases of correlated magnetic materials \cite{Sohn2017,Son2019}. In addition, it was suggested that the spin-phonon interaction can enhance the phonon amplitude of Ca\textsubscript{2}RuO\textsubscript{4} \cite{Lee2019}. The phonon anomalies of Sr\textsubscript{2}RhO\textsubscript{4} also can be induced by magnetism, such as the short-range antiferromagnetic ordering suggested by the fact that magnetic susceptibility presents a peak at 200 K \cite{Subramanian1994,Shimura1992}. However, we cannot rule out other possibilities of electron-phonon or phonon-phonon interactions for the anomalous behaviors of the Sr\textsubscript{2}RhO\textsubscript{4} phonon.\\
\forceindent In summary, we investigate non-equilibrium dynamics of photoinduced carriers in Sr\textsubscript{2}RhO\textsubscript{4} by using optical pump-probe spectroscopy. The relaxation dynamics after optical photo-excitations present clear \textit{T}-dependent anomalies across \textit{T}\textsubscript{S}$\sim$160 K. The coherent \textit{A}\textsubscript{1g} phonon oscillation changes from the sin-type oscillations to the smaller cos-type oscillations as temperature increases across \textit{T}\textsubscript{S}. We suggest that the unexpected variation in the \textit{A}\textsubscript{1g} phonon oscillations as well as in the relaxation dynamics are driven by the structural deformation related to the octahedral rotations across \textit{T}\textsubscript{S}.

\begin{acknowledgements}
This work was supported by the Institute for Basic Science (IBS) in Korea (Grant No. IBS-R009-D1 and IBS-R009-G2). K.W.K. was supported by the Basic Science Research Program through the National Research Foundation of Korea (NRF) funded by the Ministry of Science, ICT and Future Planning (NRF-2015R1A2A1A10056200 and 2017R1A4A1015564).
\end{acknowledgements}


\begin{references}
\bibitem{scfit} Considering photo-induced reflectivity change as a single exponential function $\Delta{R}/R(t) = A \exp{(-\gamma{t})}$, the slope of a linear fitting on logarithmic data should corresponds to scattering rate of the exponential decay: $ln(\Delta{R}/R) = -\gamma{t} + ln(A)$.
\end{references}
\end{document}